\documentclass[twocolumns]{aa}
\usepackage[dvips]{color}
\usepackage[latin1]{inputenc}
\usepackage{graphicx}
\usepackage{natbib}
\usepackage[english]{babel}
\usepackage{amsmath}
\usepackage{amsfonts}
\usepackage{amssymb}
\usepackage{makeidx}

\begin{document}

\title{Super-diffusion versus competitive advection: a simulation}

\author{D.~Del~Moro\inst{1}, F.~Giannattasio\inst{1}, F.~Berrilli\inst{1}, G.~Consolini\inst{2}, F.~Lepreti\inst{3,4}, M.~Go\v{s}i\'{c}\inst{5}}

\institute{$^{1}$ Dipartimento di Fisica, Universit\`a degli Studi di Roma ``Tor Vergata'', Via della Ricerca Scientifica 1, I-00133 Roma, Italy\\
$^{2}$INAF - Istituto di Astrofisica e Planetologia Spaziali, Via del Fosso del Cavaliere 10, I-00133 Roma, Italy\\
$^{3}$Dipartimento di Fisica, Universit\`a della Calabria, Ponte P. Bucci 31/C, I-87036 Rende, Italy\\
$^{4}$CNISM, Unit\`a di Ricerca di Cosenza, Ponte P. Bucci 31/C, I-87036 Rende, Italy\\
$^{5}$Instituto de Astrof\'{\i}sica de Andaluc\'{\i}a (CSIC), Apdo.\  3004, 18080 Granada, Spain}

\offprints{delmoro@roma2.infn.it}
\date{}

\abstract
{
Magnetic element tracking is often used to study the transport and diffusion of the magnetic field on the solar photosphere.
From the analysis of the displacement spectrum of these tracers, it has been recently agreed that a regime of super-diffusivity dominates the solar surface. 
Quite habitually this result is discussed in the framework of fully developed turbulence.}
{
But the debate whether the super-diffusivity is generated by a turbulent dispersion process, by the advection due to the convective pattern, or by even another process, is still open, as is the question about the amount of diffusivity at the scales relevant to the local dynamo process.}
{To understand how such peculiar diffusion in the solar atmosphere takes places, we compared the results from two different data-sets (ground-based and space-borne) and developed a simulation of passive tracers advection by the deformation of a Voronoi network.}
{
The displacement spectra of the magnetic elements obtained by the data-sets are consistent in retrieving a super-diffusive regime for the solar photosphere, but the simulation also shows a super-diffusive displacement spectrum: its competitive advection process can reproduce the signature of super-diffusion.
}
{Therefore, it is not necessary to hypothesize a totally developed turbulence regime to explain the motion of the magnetic elements on the solar surface.
}
\keywords{convection / hydrodynamics / turbulence/ Sun: photosphere}

\authorrunning{D. Del Moro et al.}
\titlerunning{Super-diffusion vs competitive advection}
\maketitle

\section{Introduction}
\label{intro}
\noindent
The outermost shell of the Sun is dominated by turbulent convection, which is the main driver of heat and entropy exchange between the different layers.
The present analytical models of convective turbulence \citep[namely, the Mixing Length Theory and the Full Spectrum Theory - ][]{MLT,CM91,CM96} are not able to describe appropriately the convective shell from the small subgranular scales ($\lesssim 100~km$) to the global scale.
On the other hand, the fully numerical aproaches \citep[e.g.,][]{Beeck2012}, while succesfull in reproducing limited regions of the Sun, are also unable to provide a complete view, due to the enormous range of temporal and spatial scales needed to simulate the whole solar convective envelope and the limited computing power available.\\
\noindent
On the observational side, a viable way to study the convective flows on the solar surface are the small scale magnetic field elements, since they are ubiquitous \citep{Romano2012,Keys2014} and they \citep[or their proxies, see: ][]{Steiner2001,Uit2006} are often used as tracers of the plasma flows \citep[e.g.,][]{Bonet2008, Berrilli2013,Berrilli2014}.
Moreover, the study of the dynamics of the magnetic elements on the solar photosphere is of particular interest in astrophysics, since it could also provide constraints on:\\
a) the characteristic spatiotemporal scales of the emergence and diffusion processes of the magnetic field \citep[e.g.,][]{Orozco2012a,Stangalini2014}\\
b) the rates of interaction between the magnetic elements themselves, which can cause magnetic reconnections and subsequent micro-flaring events: a plausible mechanism for the heating of the upper solar atmosphere  \citep[as, for example, in the model of ][]{Viticchie2006}.\\

\noindent
%
%
In order to simplify the description of the motion of the magnetic elements, we neglect the force they exert on the surrounding plasma \citep{Petrovay1994, Petrovay2001}, thus considering them as passive tracers of the underlying flows.
This assumption has been verified in the quiet Sun \citep{Viticchie2010,Orozco2012b,Bellot2012,Giannattasio2013}, where high values of $\beta$ are observed.\\
\noindent
In a Lagrangian approach, the trajectories of single magnetic elements are
employed for a statistical determination of the dynamical properties of the ensemble.
In particular, the magnetic element motion is described via the mean square displacement and modeled with a power law: $\langle\delta_r^2\rangle = k\cdot t^{\gamma}$, where $\langle\delta_r^2\rangle$ is the mean square displacement of the magnetic elements from their initial ($t=0$) position, $k$ plays the role of a diffusion coefficient, $t$ is the time span from the magnetic element first detection, and $\gamma$ is the so-called spectral index of the power law.
For diffusive motions, if $\gamma=$1, the diffusion is termed normal or Fickian, \citep[as, e.g., in the case of the Random Walk scenario - ][]{Einstein1905}.
When $\gamma\neq$1 the diffusion is termed anomalous.
Sub-diffusion if $\gamma<$1 (the diffusion process is ``slower'' than normal diffusion); super-diffusion if $\gamma>$1 (the process is ``more rapid'' than normal diffusion).
%
%
It is worth to note that only in the $\gamma=1$ case the constant $k$ coincides with the diffusion coefficient.
In the anomalous diffusion cases, the diffusion coefficient, defined as $k(\delta r,t)=\langle\delta_r^2\rangle /t$, depends on the spatial and temporal scale \citep[e.g.,][]{Schrijver96, Chae08, Abramenko11, Lepreti12}.\\
%
%
%
The diffusion of magnetic elements has been studied by several authors through segmentation and tracking algorithms of photosperic data \citep[see][ and references therein]{Goode12}.\\
The magnetic elements are detected using either the associated polarization \citep[e.g.,][]{Viticchie2009} or their excess brightness in the G-band \citep[e.g.,][]{Sanchez2010}; then, most authors utilize a Lagrangian approach to determine the spectral index $\gamma$ \citep[see, e.g.,][]{Ruzmaikin96, Utz10, Kiti12}.
See for example \cite{Jafa2014} for the most recent review on the results obtained in previous works.\\
Apart from few older reports \citep{Schrijver90,Lawrence1993,Cadavid1998,Cadavid1999}, all the works on magnetic elements diffusion agree on retrieving a super-diffusive behaviour \citep[e.g.,][]{Wang93,Schrijver96,Berger98,Hagenaar99,Chae08,Abramenko11,Chitta2012,Giannattasio2013, Giannattasio2014,Jafa2014} at all accessible spatio-temporal scales, ranging from the granulation \citep[diameter $\simeq 1$Mm and lifetime $\simeq$ few minutes; ][]{Muller99, Hirzberger99, Delmoro2004a} to the super-granulation \citep[diameter $\simeq 30$Mm and lifetime $\simeq$ a day; ][]{Raju99, Srikanth00, Delmoro2004b}.\\
In the literature \citep[e.g.,][]{Abramenko11}, $\langle\delta_r^2\rangle$ is often called ``average squared displacement'' or ``squared displacement spectrum'' or simply ``displacement spectrum''.
%
In the following, for the sake of clarity and to be consistent with previous literature we will use the term displacement spectrum when referring to the $\langle\delta_r^2\rangle$ computed from the magnetic element tracking.\\
\noindent
In this work, we used an advection simulation to reproduce the scaling laws observed in the magnetic element motion on the solar photsphere.
We assume that the plasma velocity structures (i.e.: the granules) evolve under a simple mutual repulsion model, and that the dynamics on the granular scale is completely determined by the deformation of such granular structures.
We consider the connected down-flows on the solar photosphere as a sort of lattice-like network, and suppose the magnetic elements to be bonded to the photospheric velocity sinks located at the lattice junctions (the intergranular lanes vertexes).
Under such hypotheses, the simulation returns a ``super-diffusive'' regime with a scaling similar to that observed on the solar surface.\\

\noindent
Therefore, a signal of super-diffusion can be imitated assuming that the magnetic elements are passively advected by the plasma down-flow sinks and that the evolution of the granular pattern is a competitive expansion process.\\

\begin{figure}[h!]
\centering
	\includegraphics[width=0.45\textwidth]{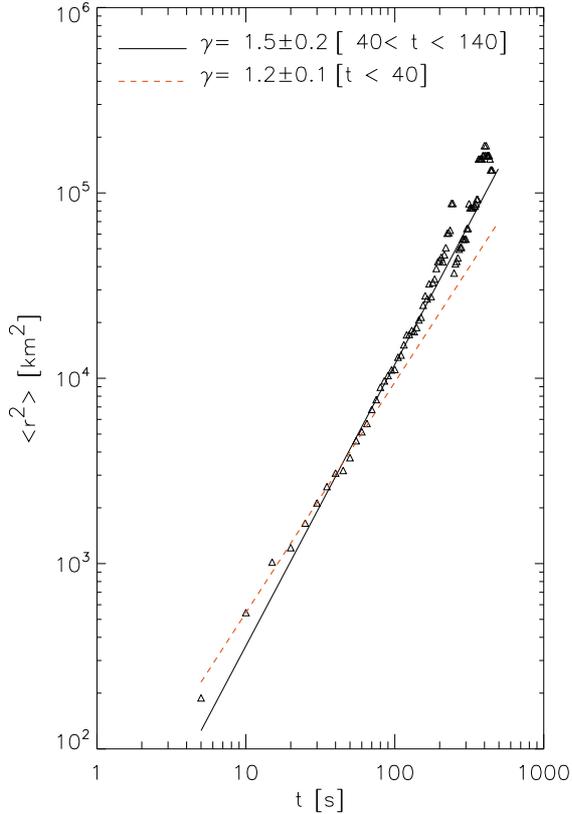}
\caption{The displacement spectrum for magnetic G-band bright points for a high-resolution data-set acquired at the NSO-DST observatory. The lines represent power law fits for the ranges $[t<40~s]$ s and $[40~s\leq t<140~s]$
.}
	\label{fig.R2_IBIS}
\end{figure}
\begin{figure}[h!]
\centering
	\includegraphics[width=0.45\textwidth]{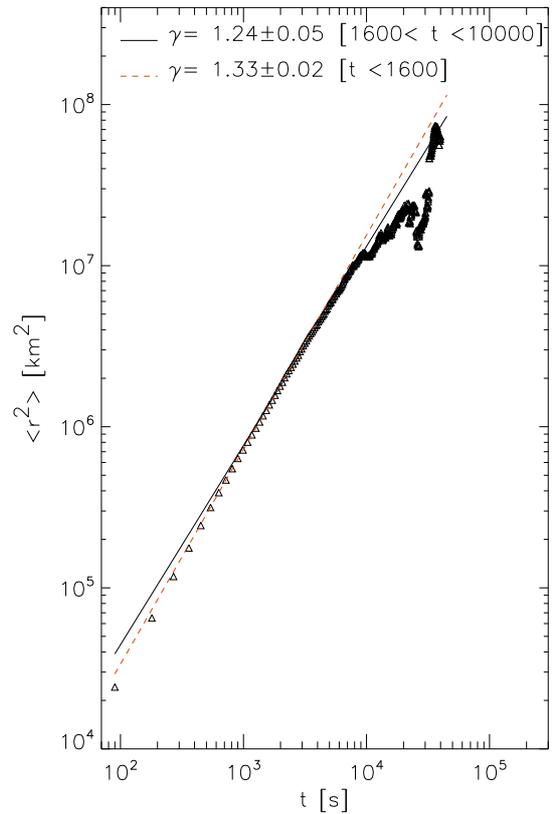}
\caption{The displacement spectrum for magnetic elements from a long-duration spectropolarimetric data-set acquired by HINODE-NFI. The lines represent power law fits for the ranges $[90~s\leq t<1600~s]$ and $[1600~s\leq t<10000~s]$.}
	\label{fig.R2_HINODE}
\end{figure}
\section{Real data}
\label{data}
\noindent
In Figs. \ref{fig.R2_IBIS} and \ref{fig.R2_HINODE}, we report the displacement spectrum $\langle\delta_r^2\rangle$ for two different data-sets imaging the quiet Sun.\\
The plot in Fig. \ref{fig.R2_IBIS} has been realized computing the displacement of G-band bright points co-spatial with magnetic field, as imaged by the ground based instrument IBIS \citep[Interferometric BIdimensional Spectropolarimeter - ][]{Cavallini2001, Cavallini2006, Viticchie2009}, located in the Dunn Solar Telescope of the National Solar Observatory (NM-USA).
This analysis is similar to that in \citet{Abramenko11}, the main difference being the check of the magnetic origin of the G-band bright points allowed by the co-temporal magnetograms.\\
The spatial resolution of this data-set, restored with MFBD \citep[Multi-Frame Blind Deconvolution - ][]{vannoort2006}, is $\simeq 0".1$, and the time between consecutive frames is $dt=5$ s.
The data have been $k_h-\omega$ filtered with a $7~km/s$ cut-off velocity.
Since data are affected by seeing, we tried to compensate this effect by applying a jitter removal algorithm.
We subtracted from each bright point displacement, a tip-tilt contribution.
Such contribution was estimated via the correlation of a $3".3 \times 3".3$ sub-frame centered on the bright point at two consecutive frames.
\\
The plot in Fig. \ref{fig.R2_HINODE} has been realized computing the displacement of magnetic elements from the long-duration ($\sim~25$ hr) data-set acquired by HINODE-NFI \citep{Kosugi2007, Tsuneta2008} and used in the analysis by \cite{Giannattasio2013,Giannattasio2014}.
The spatial resolution of this data-set is $\simeq 0".3$ and the time between consecutive frames is $dt=90$ s, also these data have been $k_h-\omega$ filtered \citep[for further details on the calibration see][]{Gosic2012}.\\
The tracking algorithm used for the two data-sets is the same used in \citet{Giannattasio2013,Giannattasio2014} and presented in \citet{Delmoro2004a} and \citet{Berrilli2005}\\
These two data-sets combined allow us to explore more consistently the time range from $5~s$ to $\simeq 10000~s$.
They consistently identify a super-diffusive behaviour ($\gamma \simeq 1.4$) in the overlapping time range $90~s\leq t<500~s$.
We stress that, to our knowledge, this is the first time that two data-sets acquired with rather different instruments and at different sites, agree on retrieving $\gamma$ values consistent within the errors.
\section{Data Interpretation}
\label{Interpretation}
\noindent
\begin{table}[h!]
\caption{$\gamma$ values retrieved by power law fits form the displacement spectra shown in Figs. \ref{fig.R2_IBIS} and \ref{fig.R2_HINODE} in different space and time ranges.}
\centering
\begin{tabular}{ccc}
\rule[-1ex]{0pt}{2.5ex} Space Range [$Mm$] & Time Range [s] & $\gamma$ value \\ 
\hline 
\rule[-1ex]{0pt}{2.5ex} $<0.05$ & $<40$ & $\simeq 1.2$ \\ 
\rule[-1ex]{0pt}{2.5ex} $0.05-1.50$ & $40-1600$ & $\simeq 1.4$ \\ 
\rule[-1ex]{0pt}{2.5ex} $>1.50$ & $>1600 s$ & $\simeq 1.2$ \\ 
\end{tabular} 
	\label{gammas}
\end{table}
On the light of these two plots, whose results are summarized in Table \ref{gammas},  we put forward a possible interpretation.\\
For times shorter than $40~s$ (and hence for space scales smaller than $\sim 50~km$), the displacement behaviour looks slightly super-diffusive: $\gamma \simeq 1.2$.
Since that time scale is not explored by the {\itshape Hinode} data-set, we cannot exclude a terrestrial atmosphere effect: despite the MFBD and the jitter correction, the G-band bright points may undergo deformations and displacements as a result of still uncorrected seeing aberrations. The seeing short ($\ll 1~s$) correlation times could affect the results.
On the other hand, the G-band bright points are not perfect passive tracers: they have a structure by their own, which can be deformed by the surrounding plasma flows.
More, buffeting-induced transverse oscillations as those found by \citet{Stangalini2013} could generate a signal likely to be detected as a movement of the tracer.
Those effect could mimick a random walk behaviour of the structure centre, for spatial scales $\lesssim 100~km$.
The third point we raise is that the short time range $t<40~s$ is lower than the correlation time-scale of granulation, suggesting that, at least on that short scales, $\gamma \simeq 1.2$ can really be the signature of anomalous diffusion by the intergranular turbulent flow field.\\
What happens for scales longer than $\sim 40$ s and larger than $\sim 100~km$? Why $\gamma$ increases and indicates a clear super-diffusive regime?\\
%
%
At those spatial and time scales all the points raised above do not hold any more, and the magnetic elements behave like passive tracers on the large scale velocity field.
Therefore, we interpret the $\gamma > 1$ regime as a real signal of super-diffusivity.
Usually, this is explicated as the result of a turbulent dispersion process and analyzed in the framework of a fully developed turbulence.
Instead, here we put forward the hypothesis that such an effect could be mainly due to the deformation of the lattice-like downflow network.
Therefore, the displacement spectrum would  also embed the contribution of the deformation of the network, which could be dominant at the large scales.
To test this hypothesis, we developed a simplified computational model to simulate such a regime.
\section{Simulation}
\label{simul}
To test our hypothesis and investigate the possible contribution of an advecting granular field in diffusing the tracers, we developed a simple simulation and studied the motion of the tracers embedded in.\\
Since the tracers used in the diffusion studies are often (if not always) embedded in the downflow network, we developed an algorithm mimicking the advective flow generated by a granular field similar to the solar photosphere.
In this simulation, the tracers are the vertexes of a Voronoi Tessellation generated from points representing the granule upflow centres.
Such vertexes are simply advected by the flow generated by the granules' evolution, and no diffusive term is represented in the simulation.
The interaction between the upflows of the granules is simulated by a repulsive force acting between the centres as a function of the associated upflow intensity.
Stronger upflows tend to push away and 'squeeze' weaker ones, causing an evolution of the Voronoi tessellation, which in this simulation represent the lattice-like downflow network.
To take into account the granules which are pushed outside of the simulation box or are 'squezeed to death' by their neighbours (a Sink term), we introduced a weak Source term as the generation in a random position of new upflows to maintain the number of granules constant.
The motion of the tracers therefore is defined only by the deformation of the lattice-like pattern specified by the upflows' position and their relative intensity.\\
%
\begin{figure}[h!]
\centering
	\includegraphics[width=0.45\textwidth]{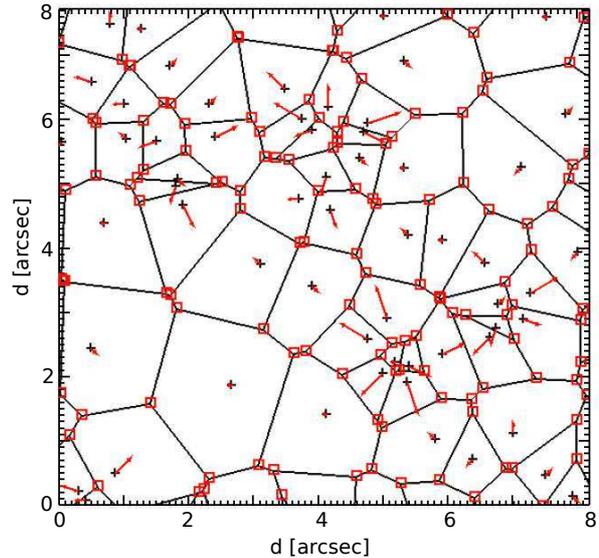}
\caption{A snapshot from the simulation. The black crosses represent the advection centres; the red vectors are the displacement ($5\times$ exaggeration) to be applied to the centres due to the neighbours' action; the red boxes on the Voronoi Tesselation highlight the cell vertexes, which are used as tracers to compute the displacement spectrum. For the sake of visualization, only about a twentieth of the simulation domain is shown.}
	\label{fig.simul}
\end{figure}
%
\noindent
The rules of the simulation are the following:\\
1) At $t=0$, $N=1024$ points are randomly distribuited on a $2048\times2048$ matrix, such that the average distance between them is $64$ pixels.\\
2) Each point $p$ has an associated interaction range $R(p)$ expressed in pixels.
At $t=0$, we initialize randomly $R(p)$ with a Gaussian distribution centred at $64$ pixels and with a standard deviation of $6$ pixels.\\
3) Every cycle, any point repels away the other points within its $R(p)$ by $1~pixel$.\\
4) At the end of the cycle, each point has its $R(p)$ decreased by the numbers of displacement it has undergone during the cycle due to the repulsion of the other points.\\
5) If a point is displaced out of the simulation bounds or its $R(p)$ falls below $1$, it is replaced by another point in a random position and with $R(p)=64 \pm 6$.\\
6) At the beginning of any cycle, all the $R(p)$ are incremented by a connectivity factor $C=4$, to compensate for the displacements caused by $4$ neighbours, on average.\\
7) The algorithm is iterated for $1000$ cycles.\\
With the suitable choice of time-step ($dt=15$ s) and pixel-scale ($px=1/64~arcsec$), this algorithm simulates the interaction between the upflows of the solar granulation.
The granules in a underdense location will expand, growing a larger and larger $R(p)$, while granules in a more crowded location may shrink and disappear, due to the other granules' pressure.
After some iterations, the simulation reproduces a statistically steady, renovating granular pattern (a sample snapshot is shown in Fig. \ref{fig.simul}).
We chose $N$, $R(p)_{t=0}=[64\pm6]$ and $C=4$ to minimize this transient phase and to obtain an average granular size of $\sim1~arcsec = 750~km$ on the solar surface.
The choice of such values, corresponds to assume that, on average, any granule interacts only with its first neighbours, since the average distance between centres is equal to the mean interaction range.\\
After the transient phase, on average, $\sim15$ granule centres have to be generated randomly in every cycle to compensate those granules which moved out of the computation bounds or were suppressed.
Thus, the simulation contains a weak Sink/Source term, since the loss or the injection of a granule centre changes the network pattern locally and cancels out/generates new Voronoi vertexes and therefore new tracers.\\
To be as consistent as possible with the real data analysis, we applied to the vertexes the same tracking algorithm used for the two data-set of Figs. \ref{fig.R2_IBIS} and \ref{fig.R2_HINODE}.\\
\begin{figure}[h!]
\centering
	\includegraphics[width=0.45\textwidth]{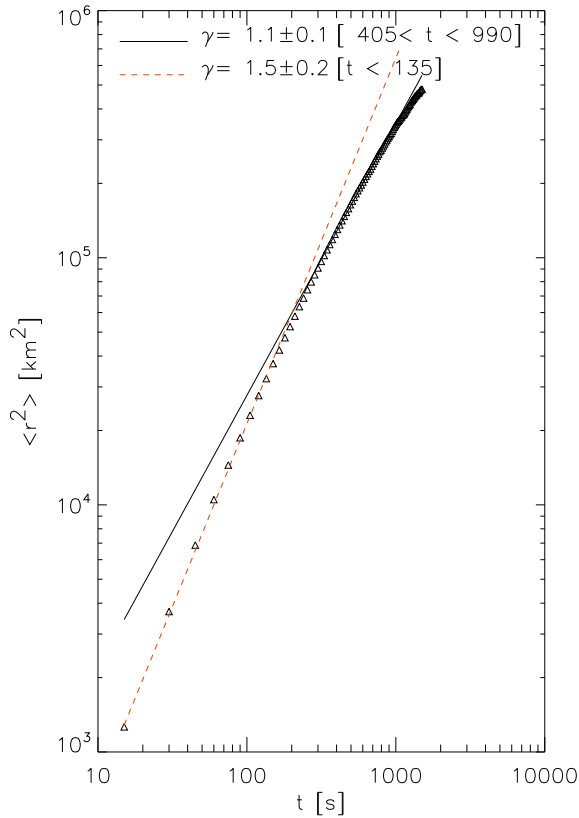}
\caption{The displacement spectrum for the Voronoi Tessellation vertexes of the advection simulation. The lines represent power law fits for the ranges $[t < 135~s]$ and $[405~s\leq t<1000~s]$.}
	\label{fig.R2_VOR}
\end{figure}
\noindent
We run several simulations, using different parameter sets, obtaining consistently super-diffusive $\gamma$ values, which are compatible with those retrieved by the observations \citep[][]{Giannattasio2013}.
The displacement spectrum retrieved by a sample simulation is shown in Fig. \ref{fig.R2_VOR}.
The exponent retrieved by a power law fit for times shorter than $135~s$ is $\gamma\simeq1.5$.
For times longer than $400~s$, the $\gamma$ value decrease to $\simeq1.1$.\\
The transition point between the two behaviours has been estimated via a fitting procedure to be $200 \pm 90 s$.
The errors on the $\gamma$ values reported in the figure represent the variability of the fit results due to changes in the fitting ranges and not the fitting errors themselves, which are tipically an order of magnitude smaller.
Changing the timestep $dt$ or the pixel scale has no effect on the index $\gamma$.
Changing instead the number of granules $N$ or the connectivity $C$ or the shape of the distribution of $R(p)$ at $t=0$, affects the average granule size and the duration of the transient phase.
The $\gamma$ values themselves depend weakly on the parameters $N$, $R(p)$ and $C$, which define the ratio between the average interaction range and the average granule dimension.
The changes of the $\gamma$ values due to small variations of the parameter sets are included in the reported errors.
Instead, those parameters affect the position of the transition between the $\gamma\simeq1.5$ and $\gamma\simeq1.1$ regimes, since they set the spatio-temporal scale over which the flow can be considered isotropic: for a scale larger than average granule dimension and longer than the average granule duration.
%
\section{Simulation Discussion}
\label{Discuss}
\noindent
The motion of the vertexes of a lattice-like pattern, undergoing deformation due to the mutual repultion of the generating points, has a displacement spectrum with a spectral index $\gamma\simeq 1.5$.
Given the parameters chosen for the simulation, the pattern produces a behaviour consistent with the solar granulation: it shows cells that expand, collapse and shuffle with characteristic time, velocity and dimension close to the granular ones.
The displacement spectrum computed from the motion of the ``downflow'' vertexes of this simulation is totally defined by the pattern evolution, which has a single stochastic ingredient: the replacement of the removed upflows with new ones in random positions.
Since we introduce only $\sim10$ new upflows on the $\sim10^3$ existent upflows every step, we estimate this should be a small contribution.
We recall that the number of new upflows introduced at each iteration is a result of the Voronoi dynamics of the simulation (not a chosen parameter), which is determined by the set of the parameters discussed above.
Therefore, the $\gamma$ value obtained should derive only from the mutual interaction of the cells, which is nor entirely stochastic (this would lead to Brownian motion and $\gamma=1$), neither ballistic ($\gamma=2$).
%
\section{Conclusions}
\label{Conc}
Several works in the literature utilized a Lagrangian approach to determine the spectral index $\gamma$ in the displacement spectra $\langle\delta_r^2\rangle \sim  t^{\gamma}$ of passive tracers on the solar photosphere \citep[e.g.,][]{Ruzmaikin96, Utz10, Abramenko11, Kiti12}.
The most recent among these works \citep[see][ and references therein]{Jafa2014}, agree in retrieving a $\gamma>1$ value, suggesting that a super-diffusion regime is present in the solar photosphere.\\
In this paper, we computed with the same algorithm the displacement spectra of magnetic elements from two spectropolarimetric data-sets: one from a ground based instrument (IBIS at NSO/DST) and the other from a space-borne instrument (NFI on-board HINODE satellite).
We confirm the superdiffusive motion of these elements, but put forward an alternative interpretation in the term of the deformation of the lattice-like structure advecting the tracers, instead of a  fully developed turbulence framework used so-far.\\
To substantiate this interpretation, we realised a simple simulation of an advection process mimicking the granular flows on the solar surface and used the vertexes of the simulated granular cell as tracers.
In this simulation, the motion of the tracers is entirely determined by the deformation of the lattice-like downflow network modeling the solar surface flow patterns.
When analysed with the same approach as the real data-sets, the simulation showed a displacement spectrum with a ``super-diffusive'' ($\gamma\simeq1.5$) behaviour.\\
This result confirms that other theoretical frameworks may be used to interpret the motion of the magnetic elements on the solar surface, and that the diffusion parameters which are calculated from the displacement spectrum analysis, may not be necessarily related to the effects of a turbulent regime.\\
For example, \citet{Escande07} investigated theoretically some cases when the $\vec{v}$ term in the advection-diffusion equation cannot be zeroed by changing reference frame (the Lagrangian approach): in the hypothesis of an evolving advection field, the tracers can jump from one cell to the other and enter into long flights, depending on the characteristic temporal, spatial and intensity scales of the advection field.
Such a point of view is definitely not far from the simple simulation we put forward.\\
Parallely, \citet{Solomon94} explored experimentally different regimes, from quasi-periodic to chaotic and to turbulent regimes, finding that both the quasi-periodic and the chaotic regimes are characterised by L\'evy-flights and show super-diffusion, with $\gamma\simeq1.5$.\\
Other analogies between the views of \citet{Escande07} and \citet{Solomon94} and our algorithm and the possibility to further extend their results \citep[possibly, applying the pair separation approach of ][]{Giannattasio2014b}, are at the moment under investigation.
%
%
%
%
%
%

%
\acknowledgements
This work is partially supported by a PhD grant at University of Rome ``Tor Vergata'' and by the Italian MIUR-PRIN grant 2012P2HRCR on ``The active Sun and its effects on Space and Earth climate'' and by the Space Weather Italian COmmunity (SWICO) Research Program..\\
Financial support by the Spanish Ministerio de Econom\'{\i}a y Competitividad through project AYA2012-39636-C06-05 (including a percentage from European FEDER funds) is gratefully acknowledged.\\
Part of the data used in this work were acquired in the framework of {\itshape Hinode} Operation Plan 151, entitled {\itshape Flux replacement in the solar network and internetwork}.
{\itshape Hinode} is a Japanese mission developed and launched by ISAS/JAXA, collaborating with NAOJ as a domestic partner, NASA and STFC (UK) as international partners.
Scientific operation of the Hinode mission is conducted by the Hinode science team organized at ISAS/JAXA.
This team mainly consists of scientists from institutes in the partner countries.
Support for the post-launch operation is provided by JAXA and NAOJ (Japan), STFC (U.K.), NASA, ESA, and NSC (Norway).\\
Part of the data used in this work were acquired with IBIS at NSO/DST under the proposal ID T748.
NSO is operated by the Association of Universities for Research in Astronomy (AURA), Inc. under cooperative agreement with the National Science Foundation.

{}

\end{document}